\documentclass[twocolumn,showpacs,preprintnumbers,amsmath,amssymb]{revtex4}

\usepackage{graphicx}
\usepackage{dcolumn}
\usepackage{bm}

\begin{document}

\preprint{APS/123-QED}

\title{Measurement and analysis of the Hall effect of A-Fe$_2$As$_2$  single crystals \\
with A = Ba, Ca or Sr.}

\author{C. L. Zentile}%

\author{J. Gillett}%
\author{S. E. Sebastian}%
\author{J. R. Cooper}
\affiliation{Cavendish Laboratory, University of Cambridge, J. J. Thomson Avenue, Cambridge, CB3 0HE, United Kingdom}

\date{\today}

\begin{abstract}
We report measurements of the Hall coefficient $R_H$ for single crystals of
 AFe$_2$As$_2$ with $A = Ba, Ca$  or $Sr$ which are the anti-ferromagnetic
parent compounds of some high temperature pnictide superconductors. We show that $R_H$
of Sr-122 is consistent with high field quantum oscillation data.  Our $R_H(T)$ data
can also be used to estimate values of the spin density wave gap, giving
$\Delta_{SDW}(0) = 710\pm 70$\,K for Sr-122 and $435\pm 20$\,K for Ba-122.

\end{abstract}

\pacs{Valid PACS appear here}
\maketitle


The discovery of high temperature superconductivity in the iron pnictides has
generated considerable scientific interest. They are correlated electron systems
with quasi-2D structure,
  and show both similarities with, and notable differences from, the cuprates.  Here we report
  measurements of the Hall coefficient $R_H$ for three isostructural `122' iron
  arsenides that all  have a structural phase transition at a temperature  $T_S$.
   It is believed\,\cite{Zhao,Huang,Goldman}
  that anti-ferromagnetic (AF) order also sets in at  $T_S$, and that
  superconductivity  occurs when this is suppressed  by doping with K\,\cite{Rotter}, Na or Co
  or by
  applying pressure\,\cite{Alireza}.

\begin{figure}
\includegraphics[width=0.48\textwidth, bb=58 449 572 713]{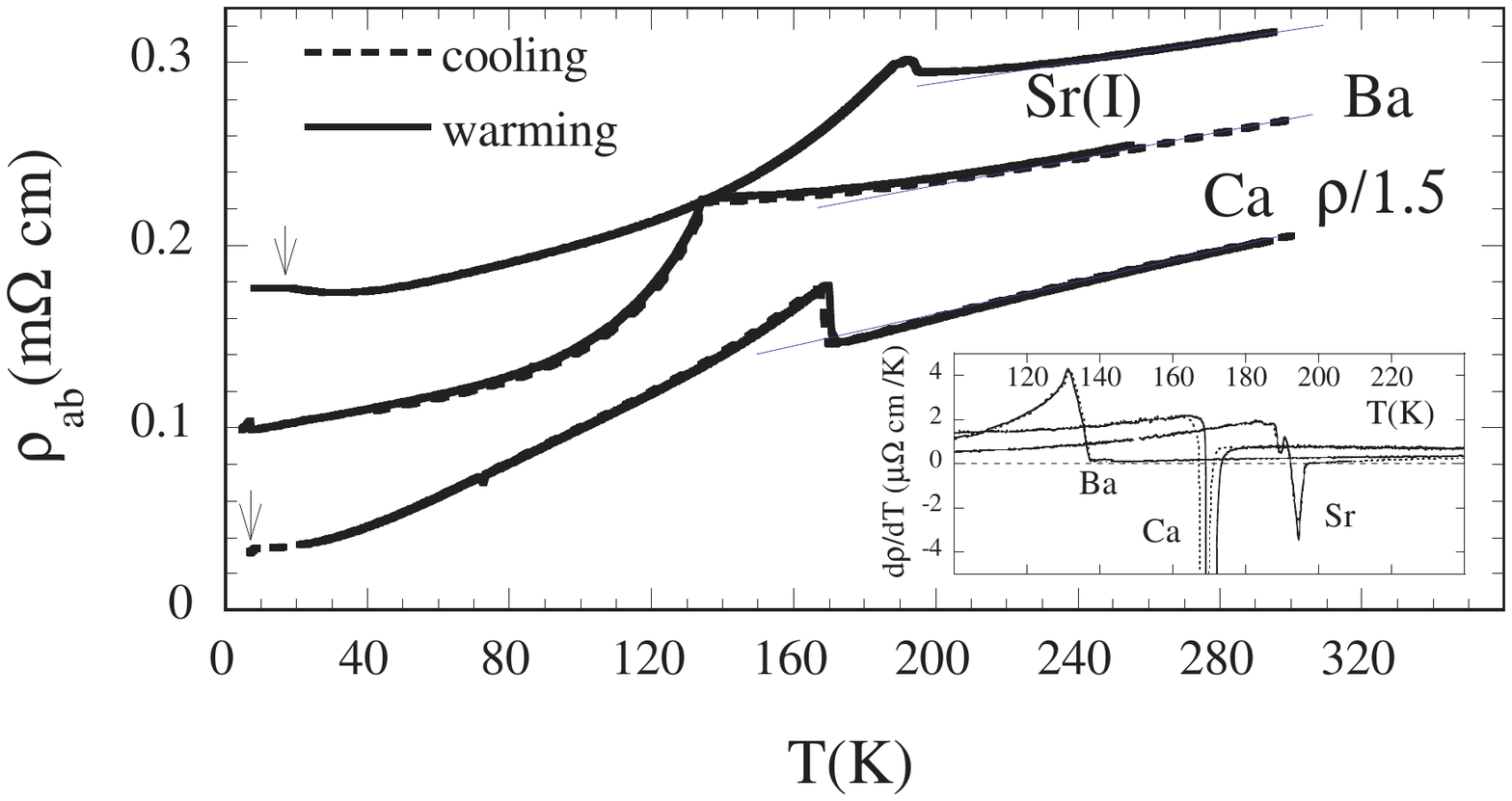}
\caption{\label{fig:Fig1} Zero-field resistivity as a function of $T$, $\rho(T)$ for
single crystals of AFe$_2$As$_2$ (A = Ba, Ca, Sr). The arrows show changes in slope
arising from a small amount of parasitic superconductivity\,\cite{Saha,Hiramatsu}. The
inset shows $d\rho/dT$
 near the phase transitions.}
\end{figure}

 Fig.\,\ref{fig:Fig1} shows the in-plane resistivity $\rho(T)$ versus $T$
 for three of the crystals for which $R_H$ was measured just over a year ago.
 While their resistivity ratios ($RR$) summarized in Table\,\ref{tab:table1} are slightly lower than
published values\,\cite{Rullier,Ronning,Chen,YanCan}, higher values have since been
obtained\,\cite{PSyers} and Sr crystals from similar growth batches were used for
quantum oscillation (QO) experiments\,\cite{Sebastian1}. Also our  Hall mobility
ratios $\mu_H(T\rightarrow 0)/\mu_H(300)\approx 200$ are substantially higher than the
$RR$ values. Further evidence for sample quality is the sharpness of the phase
transitions shown by the derivative
 plots $d \rho/dT$ in the inset to Fig.\,\ref{fig:Fig1}. The
 $T_S$, values in  Table\,\ref{tab:table1}, defined by the maxima in $|d\rho/dT|$ for
  Sr and Ca and from the temperature of the step increase in $d\rho/dT$ for Ba, agree well with
  published data\,\cite{Rullier,Ronning,Chen,YanCan}. Above $T_S$, $\rho (T)$ for
 the Ba and Sr crystals
 shows upward deviations from linearity as indicated by the faint lines in Fig.\,\ref{fig:Fig1} and
 by the fall in  $d\rho/dT$ shown in the inset. This could arise from the onset of magnetic or structural fluctuations
 $\sim 40$\,K above $T_S$. In contrast, Ca-122 shows a slight downward
 deviation from linearity above $T_S$.
 Below $T_S/2$ the $\rho(T)$ curves vary
   as $A + BT$  over a considerable range of $T$. For Sr-122 this is followed
    by a slight upturn  below 30\,K
   reminiscent of the Kondo effect
    while for Ca-122 $\rho(T)$ flattens out as $T^3$
  below 30\,K. Unusually, $\rho(T)$  for Ba-122 falls linearly to the lowest measured temperature of
5\,K.


\begin{figure}
\includegraphics[width=0.5\textwidth, bb=73 419 410 703]{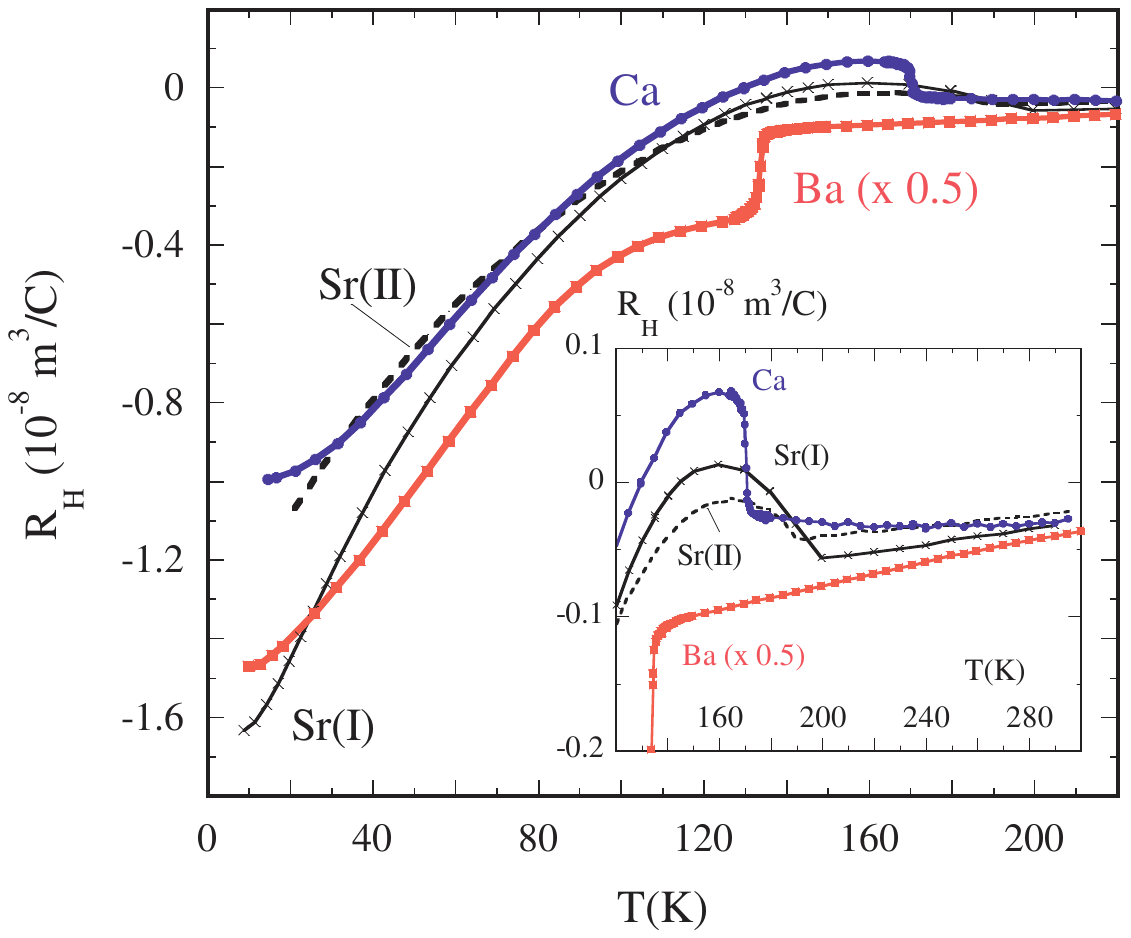}
\caption{Color online: $T$ dependence of the Hall coefficient, $R_H$(T) for single
crystals of AFe$_2$As$_2$ (A = Ba, Ca, Sr) with a field of 5\,T $\parallel c$. The
inset shows the transitions and high-$T$ region on an expanded scale.}
\label{fig:Fig2}
\end{figure}
Hall effect measurements were made on crystals  of Ca-122 and  Sr-122, from two growth
batches, all grown using Sn flux. These long, thin bars were $1.7-0.5 \times 0.6-0.3
\times 0.1-0.05$\,mm$^3$ in size and had two pairs of Hall contacts that were measured
simultaneously to check for uniformity and reproducibility. The Van der Pauw
four-contact method was used for a thin flake of Ba-122, $1.2\times 1.2\times
0.02$\,mm$^3$ in size, grown using self-flux. The magnetic field was directed along
the $c$-axis and was repeatedly reversed by  rotating the sample stage by
180\,$^\circ$. Our $R_H(T)$ curves in Fig.\,\ref{fig:Fig2} show remarkably similar
shapes, signs and
 magnitudes for the four samples.
$R_H$ increases approximately linearly
 by a factor of 3 between 20 and 60-80\,K followed by notable flattening
  well below
  $T_S$. This flattening could be associated with the proposed spin density wave
   (SDW) state. Below 120\,K $R_H$ is negative for all the samples but for Ca-122 and the
  higher purity
  Sr(I) crystal there is a  region between $T_S$ - 40 and $T_S$ where $R_H > 0$,
 suggesting a different balance between
   electron and hole contributions there.
    Above $T_S$, $|R_H|$ for
    Ca-122 rises slightly and then becomes constant. The other
    two compounds show a linear increase in $|R_H|$ between 300\,K and $T_S^+$.

Three QO studies of  in AFe$_2$As$_2$ in magnetic fields as high as 55\,T have been
reported\,\cite{ Sebastian1, Analytis, Harrison} and the results analyzed in terms of
small ellipsoidal
 pockets. We find that our Hall data are consistent with the QO results, especially for Sr-122.
\textit{A priori} they could be very different because there is an even number of
electrons per unit cell in both the high-$T$ tetragonal phase and the low-$T$ phase
where there is a commensurate anti-ferromagnetic (AF) wave
vector\,\cite{Zhao,Huang,Goldman} that halves the  in-plane Brillouin zone area. In
both cases the number densities of electrons ($n_e$) and holes ($n_h$)  will be the
same and their contributions to $R_H$ should be weighted by the corresponding
electrical conductivities $\sigma$, e.g. for two bands:

\begin{equation}
R_H = \frac{|R_{H}^{h}| \sigma_{h}^{2} - |R_{H}^{e
}|\sigma{_e}{^2}}{\left(\sigma_{e}+\sigma{_h}\right)^{2}}
\end{equation}

  So if
$n_h$ and $n_e$ are the same, and $\sigma_h \simeq \sigma_e$,  $R_H$ can be very small indeed,
and will not give a valid measure of the carrier concentration. It might well
 vary strongly from sample to sample depending on the precise balance between
  $e$ and $h$ contributions. We see from Table\,\ref{tab:table1} that there is actually
good agreement between our values of $R_H$(10\,K) and those in the literature despite
   the differences in the $RR$ values mentioned earlier.
Another possible source of discrepancies between high field QO and lower
 field Hall effect work
is magnetic breakdown (MB)\,\cite{Shoenberg}.  In a weak-coupling SDW system, the
energy gap ($\Delta_{SDW}$)  is  small\,\cite{Gruner}; $\Delta_{SDW} = 1.76
k{_B}T{_S}$ since
 the gap equation for an SDW has the same form as in BCS superconductivity.
 In practice, and perhaps surprisingly, any MB effects seem to be small.

 For a single  ellipsoidal pocket, the
standard, nearly free electron formulae $R_H= 1 / ne$ and $\sigma_i=n e^{2}\tau /
m_{i}$ are still valid\,\cite{Mott&Jones}.
 Here $m_i$ is the effective mass along
 a principal axis ($i$) of the ellipsoid, and  $n=2V_k/(2\pi)^3$, where
  $V_k$ is its volume in $k$-space.
 For Sr-122 three pockets labelled $\alpha$, $\beta$ and $\gamma$ were detected\,\cite{Sebastian1}
 with frequencies of $370\pm20$, $140\pm20$ and $70\pm 20$\,T and in-plane
 $k$ space areas of 1.38, 0.52 and $0.26\,\%$ of the paramagnetic Brillouin zone area,
 $(2\pi/a)^2$, where the in-plane lattice parameter\,\cite{Zhao,Ronning} $a = 3.93$\,\AA. Angular
  studies showed that the ellipsoids
 were elongated along the crystallographic $c$-axis  with  axis ratios ($r$) of 1.4, 6.1 and 3.3 for $\alpha$,
 $\beta$ and $\gamma$ respectively.  Later
 three similar frequencies and smaller $m^{\ast}$ values were found for Ba-122 crystals\,\cite{Analytis}.
 Band-structure calculations\,\cite{Sebastian1,Analytis} predict 4 inequivalent pockets in the AF Brillouin zone
arising from imperfect nesting of the AF vector
 $\overrightarrow{Q} = (\frac{\pi}{a},\frac{\pi}{a},0)$.
 As discussed below,  analysis
 of the Hall data suggests that the  largest electron pockets are absent for Sr-122.

From the values of $r$ and the $k$ space areas we obtain $V_k$ for each type of
ellipsoidal pocket. Each pocket  in the reduced AF Brillouin zone gives $n_\alpha =
0.0053(5)$, $n_\beta = -0.0054$ and $n_\gamma= -0.001(0)$ carriers per formula unit
($f.u.$).  Here the $+$ and $-$ signs denote $h$ and $e$ contributions to $R_H$
respectively. Within the error bars these satisfy $\Sigma n_e=\Sigma n_h$ and imposing
this as a precise  constraint does not affect our conclusions.
 Both papers\,\cite{Sebastian1, Analytis}  identify  $\alpha$
 as a ``tear-shaped'' hole pocket, and $\beta$ as an ellipsoidal electron
pocket. The small $\gamma$ pocket
is identified as electron-like for Sr-122 in Ref. \onlinecite{Sebastian1} and as hole-like for Ba-122 in Ref.
\onlinecite{Analytis}. Using the analysis described below we  get good agreement with our Hall data
if the $\gamma$ pocket is electron-like for Sr-122.

 In order to calculate $R_H$ we need
to find appropriate values of $\sigma_\alpha$, $\sigma_\beta$  and $\sigma_\gamma$ for use
 in Eqn.\,1. Experimentally\,\cite{Mackenzie} it was found that the isotropic mean free
 path approximation worked well for Sr$_2$RuO$_4$.
Here  we are dealing with small values of $k_F$ but any impurity potentials will still
be screened over short distances ($d$) because of the underlying metallic state that
is weakly perturbed by a SDW. In this limit, $k_Fd\ll 1$ we expect a $k$-independent
scattering cross-section, $4\pi d^2$\,\cite{SchiffQM}, which will indeed give a mean
free path that is independent of $k$. The   relative contributions to $\sigma$ in
 Eqn.\,1 are then given by $n_\alpha/k{_F^\alpha}$ etc. where $k{_F^\alpha}$ is the in-plane
Fermi wave-vector of the $\alpha$ pocket and give a negative value for $R_H$.
\begin{table}
\caption{Parameters obtained in the present work or used in the analysis. $T_S$
 values were obtained on cooling; for $A=Ca$ the warming transition
 is 2\,K higher.  $RR$ is the resistivity ratio  $\rho(300)/\rho(10)$.
 $R_H$ is the Hall coefficient measured at 5\,T,  $n_{eff}$ is the corresponding
  carrier concentration in a
 single band analysis. $\mu_H$ is the Hall
  mobility, $R_H \sigma$,  and $\beta$ is the $T^2$ coefficient
  of the inverse mobility  obtained from fits
   to $1/\mu_H = \alpha +\beta T^2$ between 10 and 60\,K.
   Sr(I) is from growth batch 170 and Sr(II) batch 105.}
\begin{ruledtabular}
\begin{tabular}{lcccc}
   & Ba & Ca & Sr(I) & Sr(II)\\
$V_{f.u.}$(\AA$^3$) & 102 &89  & 95 &95\\
$T_S$\,(K)& 135.4 & 168.6 & 194.6 & 190.8 \\
$RR$ &2.65&6.2&1.85&1.5\\
$RR$ (other work) &(4.3)$^a$&(11)$^b$&(4.7)$^c$&\\
$R_H$(295)(10$^{-10}$m$^3$/C)&  -7.6 & -3.0& -3.0 &-2.3 \\
$R_H$(T$_S$+5)(10$^{-10}$m$^3$/C)& -21.0& -2.5 &-5.5 &-4.1 \\
$R_H$(10)(10$^{-10}$m$^3$/C)& -292 & -99 & -162 &-120 \\
$R_H$(10)(10$^{-10}$m$^3$/C)& (-220)$^a$& (-95)$^b$&
(-135)$^c$ & -\\
$n_{eff}$(10)/f.u.& 0.022 & 0.056 &0.037 &0.049 \\
$\mu_H$(295)(cm$^2$/Vs) & 2.8 & 0.98 & 0.95 & -\\
$\mu_H$($T_S+5)$(cm$^2$/Vs) & 9.3 & 1.3 & 1.9 &- \\
$\mu_H(0)$(cm$^2$/Vs at 1T)   & 450 & 200 & 200 & 50\\
$\beta$(1\,$T$)(10$^{-6}$\,Vs/cm$^2$/K$^2$) & 0.8 & 2.9 & - & 5.3\\
$\beta$(5\,$T$)(10$^{-6}\,$Vs/cm$^2$/K$^2)$& 0.9 & 3.3 & 4.1 & 5.1\\
\end{tabular}
\end{ruledtabular}
$^a$ Ref.\,\onlinecite{Rullier},$^b$ Ref.\,\onlinecite{Ronning},
$^c$ Refs.\,\onlinecite{Chen,YanCan}
\label{tab:table1}
\end{table}

Band-structure calculations\,\cite{Sebastian1,Analytis} seem to give two pockets of
each type in the AF Brillouin zone. Taking this into account leads to an effective
electron concentration $n_{eff} = 0.039$ per $f.u.$. This is in excellent agreement
with the value given in Table\,\ref{tab:table1} for Sr-122 for a field of 5\,T at low
$T$. Although
 $\Sigma n_e = \Sigma n_h$ within error bars, there is little $e-h$ cancellation  in $R_H$  because the $\alpha$ and
$\beta$ pockets have very different eccentricities $r$, which means that
 $\sigma_\alpha \neq \sigma_\beta$.  If instead we included the fourth and
largest electron pocket in this analysis, using the eccentricity implied by the band-structure
calculations, then the calculated value of $n_{eff}= 0.11$ per
$f.u.$, which is much larger than the experimentally determined values.

 \begin{figure}
\includegraphics[width=0.5\textwidth, bb=44 413 428 712]{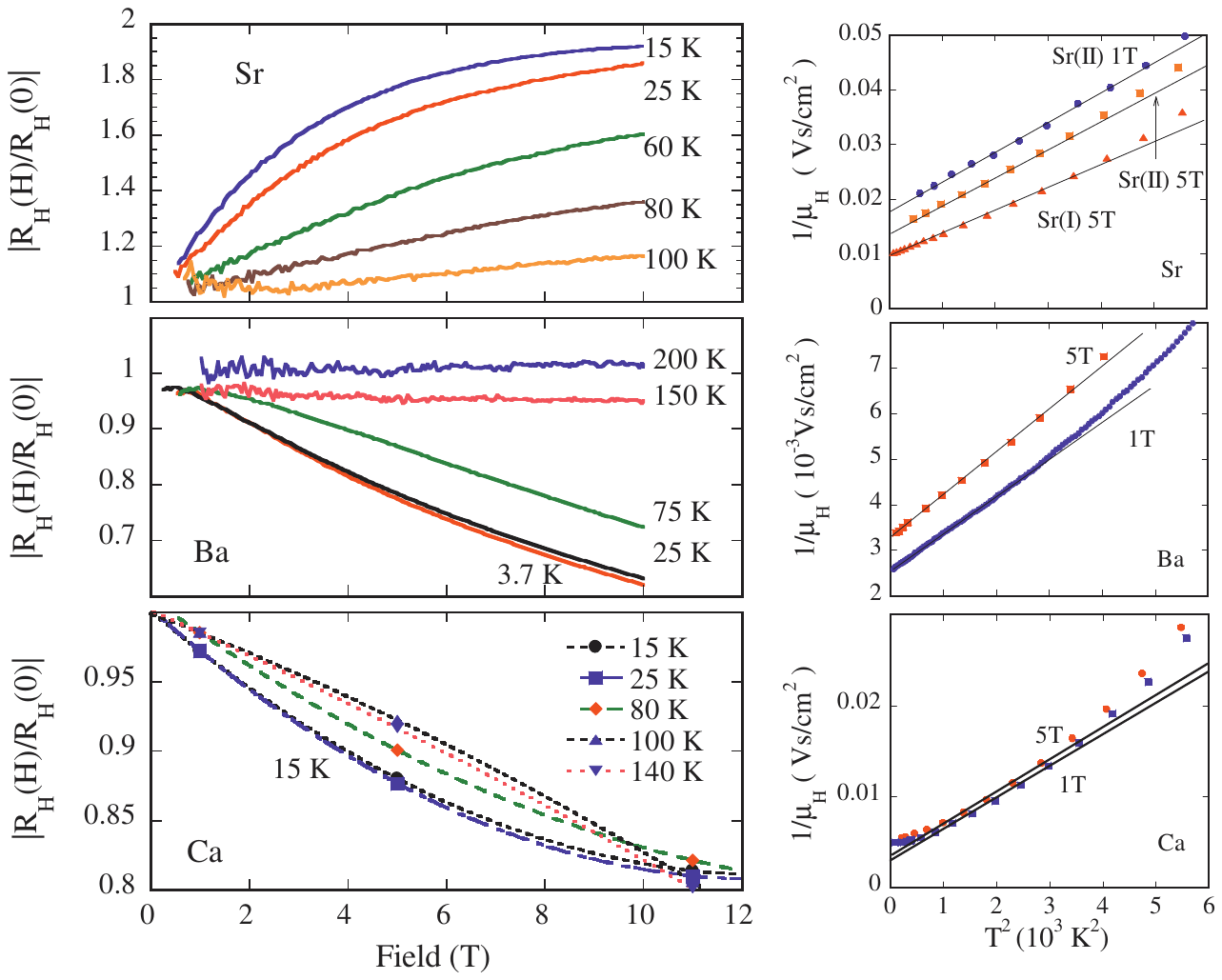}
\caption{Color online: left-hand panels, normalized magnetic field dependence of the measured Hall coefficient
 $|R_H|$ for the three compounds studied here, at several fixed $T$. Right hand panels,
  inverse Hall mobility $vs.$ $T^2$ for the 3 compounds studied.}
\label{fig:Fig3}
\end{figure}

The above analysis has all been based on $R_H$ data taken at 5\,T. As shown in
Fig.\,\ref{fig:Fig3} $R_H$ is in fact field-dependent. We do not understand this, but
although the three materials show completely different $H$- dependences, it could
still
 be a general property of the SDW state because of the $e-h$ compensation described above.
  For the Sr-122 crystal
$R_H$ increases by nearly $50 \%$ between 1 and 5\,T. In principle this could be
caused by one or more of the pockets approaching the usual high field  condition,
$\mu_H H$, i.e. $\omega_c \tau \sim 1$ at 5\,T.  But for the same crystals we did not
observe any QOs in fields from 12 to 15\,T down to 1.4\,K.  Therefore any explanation
along these lines seems to require a very small pocket of frequency $\approx$ 50\,T or
less whose period was too long to be observed in our measurements. For Ca-122 the $H$
dependence is much smaller and of opposite sign, while that for Ba-122 is larger and
also has the opposite sign to that in Sr-122.  High field orbital effects are somewhat
more likely here because Ba-122 has a higher average mobility and our Ca-122 crystal
has comparable mobility to the Sr crystal but shows a much smaller $H$-dependence.

Applying the same analysis to available data for Ba-122\,\cite{Analytis} does not give
such straightforward results. If the largest $e$-pocket is again
 absent\,\cite{Analytis}, we can satisfy $\Sigma n_e =\Sigma n_h$ and account for the
 \textit{low field} value of $R_H$ by assuming
that the $\beta$ pocket has $r$ = 3, rather than $5\pm1$\,\cite{Analytis}, that
$\gamma$ is
  in fact an $e$-pocket
and by assuming that the mean free path on the $\beta$ and $\gamma$ electron pockets
 is 20$\%$ larger than
that on the $\alpha$ pockets.  For Ca-122\,\cite{Harrison}, the two pockets observed
 have very different volumes so we have not attempted to calculate $R_H$.
 \begin{figure}
\includegraphics[width=0.4\textwidth, bb=36 114 586 638]{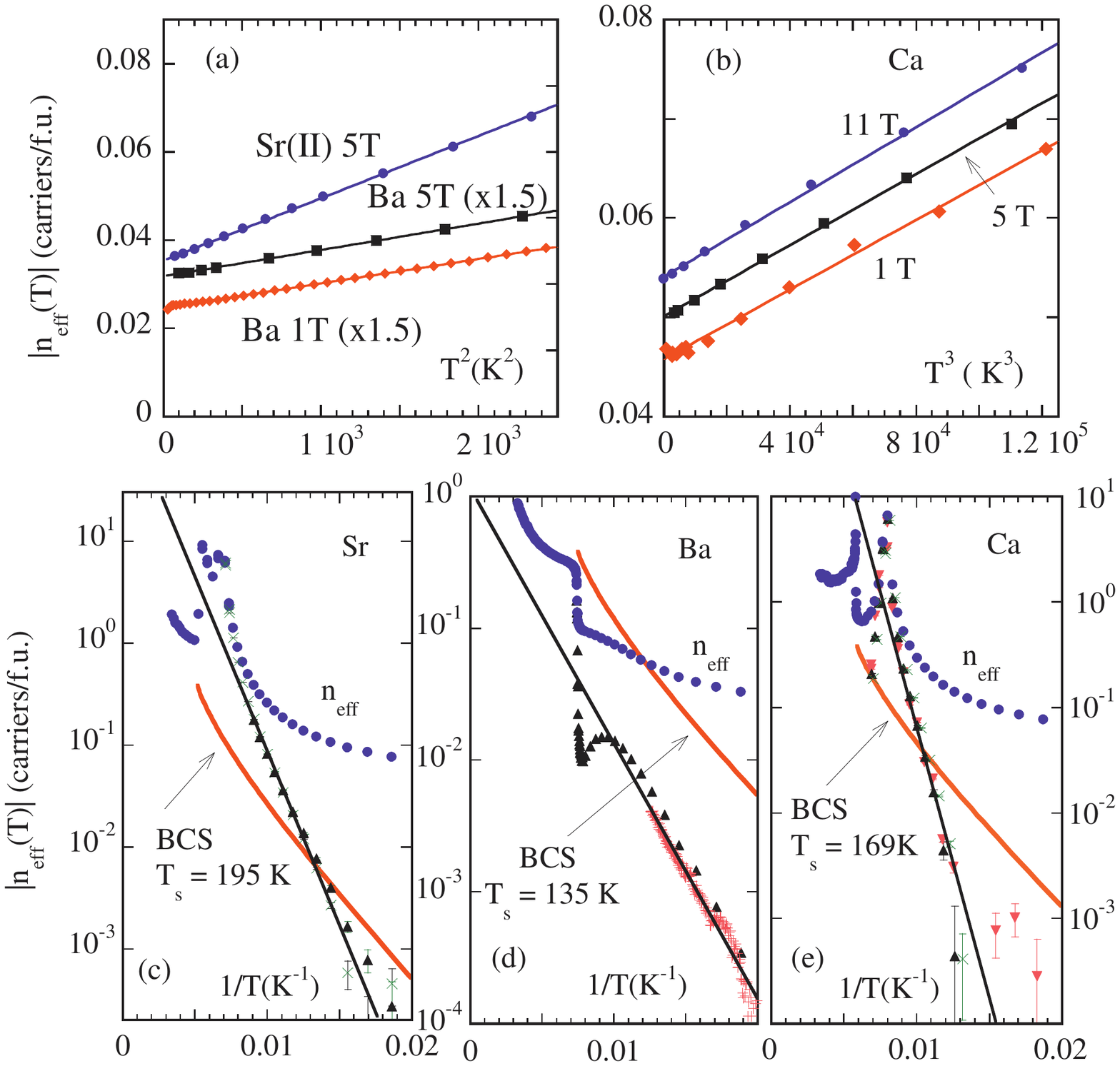}
\caption{\label{fig:Fig4}Color online:  (a) plots of effective carrier concentration
$n_{eff}$ $vs.$ $T^2$ for Sr-122 and Ba-122. The lines show the fits to $C+DT^2$. (b)
$n_{eff}$ $vs.$ $T^3$ for Ca-122 and the corresponding fits for three values of
applied magnetic field (T). (c) -(e) semi-log plots $n_{eff}(T)$ $vs.$ $ 1/T$ before
and after subtraction of the limiting low $T$ dependence. For Sr and Ba, black
 triangles show data at 5\,T while crosses show data at 1\,T.
For Ca  black triangles, inverted triangles and crosses show 5\,T, 1\,T and 11\,T data respectively.
The thicker, slightly curved, (red) lines show calculated
 weak-coupling BCS behavior for the $T_S$ values shown. The slopes of the  black lines are used to find
 the activation energy, i.e. $\Delta_{SDW}(0)$. }
\end{figure}

Fig.\,\ref{fig:Fig3} also shows that the inverse Hall mobility of all crystals studied
obeys an $\alpha$ + $\beta T^2$ law, although the range of this fit is very small for
Ca-122 and actually a $T^3$ law gives a better fit.  For the cuprate superconductors
its presence over a wide range of hole doping with little change in $\beta$ is  an
important empirical observation. In the present case, as for Co-doped
 crystals\,\cite{Rullier} it seems to arise from $e-h$ pockets whose $E_F$ values are extremely
low. From the QO data we find $E_F$ equals 240, 120 and 100\,K for the $\alpha,\beta$
and $\gamma$\,\cite{massnote} pockets in Sr-122 and 490, 403 and 115\,K for Ba-122. As
shown in  Table\,\ref{tab:table1}  the $\beta T^2$ term in $1/\mu _H(T)$ is  a factor
4 larger for Sr-122, reflecting the smaller  $E_F$ values listed above.

We now address the $T$ dependence of $R_H$ at low $T$ and propose a method of
estimating the SDW gap from our data. An important clue here is  that for Ba-122 the
electronic heat capacity above $T_S$ is $\sim8$ times larger than the low $T$
value\,\cite{Storey}. This suggests that, although the small pockets do give a finite
density of states (DOS) at $E_F$, any square-root singularities in the DOS from the
SDW will  be much larger.
 As shown in Fig.\,\ref{fig:Fig4}a, for Sr-122 and Ba-122 we can fit $n_{eff}(T)$ below 40-50\,K to a $C+DT^2$ law.
This $DT^2$ term could arise partly from   the  low values of $E_F$ combined with the
constraint $\Sigma n_e=\Sigma n_h$, leading to changes in $n_e$ or $n_h$ as $T$ is
increased.  A second possible reason is that in Eqn.\,1,
 $\sigma_{\alpha,\beta,\gamma}$ could have different $T^2$ terms from $e-e$
 scattering;
 a third is that the pockets expand as the SDW gap decreases. While the latter must
be true at higher $T$, in Fig.\,\ref{fig:Fig4} we also show   $n_{eff}$ $vs.$ $1/T$
plots calculated\,\cite{calcfootnote} for the fully nested
 case where there are no pockets and the energy gap
$2\Delta_{SDW}(T)$ has the usual  BCS $s$-wave $T$-dependence. These plots are linear
below $T _{SDW}/2$ implying that $\Delta_{SDW}$ is effectively constant and  any
expansion
 of the pockets is not important below
$T_{SDW}/2$.  As shown in  Fig.\,\ref{fig:Fig4}, subtraction of the $C + DT^2$ terms
leads to linear regions in plots of $\log n_{eff}$ $vs.$ $1/T$.   From the slopes of
these lines we
 find $\Delta_{SDW}(0)= 710\pm 70$ \,K for
Sr-122, a factor of 2.1 larger than the BCS value, and $435 \pm 20$\,K for Ba-122, a
factor of 1.8 larger than BCS.  For Ca-122 the value of $\Delta_{SDW}(0)$ is less
certain. As shown in Fig.\,4b,
 $n_{eff}$ = $C+ DT^3$ gives a good fit below 50\,K. Subtracting this gives
 $\Delta_{SDW}(0) = 1150 \pm 150$\,K,  a factor 3.9 larger than BCS. On the other
 hand if $n_{eff}$ is forced to fit $C + DT^2$ at low $T$, then a much smaller value of
 $\Delta_{SDW}(0)$ $\sim$ 1.2 times the BCS value, is obtained.

  Single crystal  optical reflectivity data
taken at 10\,K have been analyzed in terms of two broad Lorentz-Drude peaks centered
 at 360 and 890\,cm$^{-1}$ for Ba-122\,\cite{Hu,Degiorgi}
and at 500 and 1360\,cm$^{-1}$ for Sr-122\,\cite{Hu}. The lower peaks have similar
energies to our values, $\Delta_{SDW}(0)=300 \pm 15$\,cm$^{-1}$ for Ba-122 and $490
\pm50$\,cm$^{-1}$ for Sr-122. Moreover, there are clear changes in slope in the raw
reflectivity data\,\cite{Hu} at $ca.$ 270 and 600\,cm$^{-1}$ for Ba-122 and at 410 and
900\,cm$^{-1}$ for Sr-122  that are absent above $T_S$. We therefore suggest  that the
lower reflectivity anomaly could correspond to the onset of electron excitations
between the $e-h$ pockets at $E_F$ and the square root singularities from the SDW  at
$E_F \pm\Delta_{SDW}$ which would have an  energy of $\Delta_{SDW}(0)$, while the
upper anomaly could be a measure of  $2\Delta_{SDW}(0)$.

In summary, our Hall measurements on Sr-122 are surprisingly consistent with high-field quantum oscillation
data, and subject to some minor modifications, with those observed for Ba-122. We have been
able to make approximate estimates of the SDW gaps from the Hall data. For Sr and Ba we propose that these are
consistent with optical reflectivity although detailed calculations of the latter are still needed.

We are grateful to S. Battacharya, T.M. Benseman, A. Carrington, J.W. Loram, and J.G.
Storey for helpful discussions and to the EPSRC (U.K.) for financial support.


\begin{thebibliography}{99}
\bibitem{massnote} We took $m^{\ast}$ data for $\alpha$ and $\beta$ from Ref. \onlinecite{Sebastian1} and
$m^{\ast} = 0.9 m_e$ for the $\gamma$ pocket from Ref.\,\onlinecite{Sebastian2}.
\bibitem{calcfootnote} Here $n_{eff}$ is the number of electrons excited above the gap. There
will be an equal number of holes below the gap, but usually enough $e-h$ asymmetry to
give $n_{eff} \sim \exp(-\Delta/T)$.
\end{thebibliography}

\end{document}